\documentclass[12pt, draftclsnofoot, onecolumn]{IEEEtran}
\IEEEoverridecommandlockouts

\usepackage{cite,caption}
\usepackage{fleqn}
\usepackage{eufrak}
\usepackage{amsfonts}
\usepackage{amsmath,amssymb,amsfonts,amsthm}
\usepackage{algorithm,algorithmic}
\allowdisplaybreaks
\usepackage{graphicx}
\usepackage{textcomp}
\usepackage{bm}
\usepackage{pifont}
\usepackage{bbm}
\usepackage{stfloats}
\usepackage{indentfirst} 
\usepackage{hyperref}
\usepackage{color}
\usepackage{url}
\usepackage{array}
\usepackage{latexsym}
\usepackage{mathrsfs}

\ifCLASSINFOpdf
\else
\fi
\begin{document}
\title{Caching and Computation Offloading in High
	Altitude Platform Station (HAPS) Assisted
	Intelligent Transportation Systems}

\author{Qiqi~Ren,
	    Omid~Abbasi,~\IEEEmembership{Graduate Student Member,~IEEE,}
	    Gunes~Karabulut~Kurt,~\IEEEmembership{Senior~Member,~IEEE,}
        Halim~Yanikomeroglu,~\IEEEmembership{Fellow,~IEEE,}
        and~Jian~Chen,~\IEEEmembership{Member,~IEEE}% <-this % stops a space
\thanks{This work is supported in part by the National Natural Foundation of
	China (Grant No. 61771366 and Grant No. 61901312), and in part by Huawei Canada.}% <-this % stops a space
\thanks{Q. Ren and J. Chen are with the State Key Laboratory of Integrated Service Networks, Xidian University, Xi' an 710071, Shaanxi, P. R. China, (email: renqiqi5277@gmail.com; jianchen@mail.xidian.edu.cn) (Corresponding Author: Jian Chen).}% <-this % stops a space
\thanks{O. Abbasi and H. Yanikomeroglu are with the Department of Systems and Computer Engineering, Carleton University, Ottawa, ON, Canada (email: \{omidabbasi, halim\}@sce.carleton.ca) }% <-this % stops a space
\thanks{G. Karabulut Kurt is with the Department of Electrical Engineering,  Polytechnique Montr\'eal, Montr\'eal, Canada (e-mail: gunes.kurt@polymtl.ca)}}

\maketitle
\begin{abstract}
Edge intelligence, a new paradigm to accelerate artificial intelligence (AI) applications by leveraging computing resources on the network edge,  can be used to improve intelligent transportation systems (ITS). However, due to physical limitations and energy-supply constraints, the computing powers of edge equipment are usually limited. High altitude platform station (HAPS) computing can be considered to be a promising extension of edge computing. HAPS is deployed in the stratosphere to provide wide coverage and strong computational capabilities. It is suitable to coordinate terrestrial resources and store the fundamental data associated with ITS-based applications. In this work, three computing layers, i.e., vehicles, terrestrial network edges, and HAPS, are integrated to build a computation framework for ITS, where the HAPS data library stores the fundamental data needed for the applications. In addition, the caching technique is introduced for network edges to store some of the fundamental data from the HAPS so that large transmission delays can be reduced. We aim to minimize the delay of the system by optimizing computation offloading and caching decisions as well as bandwidth and computing resource allocations. The simulation results highlight the benefits of HAPS computing for mitigating delays and the significance of caching at network edges.
\end{abstract}

% \fi
\IEEEpeerreviewmaketitle

\section{Introduction}
\IEEEPARstart{T}{he} past decade has witnessed the rapid development of autonomous driving technology, which has mainly been due to progress in deep learning and artificial intelligence (AI)\cite{surveyDriving}. Autonomous decision-making systems embedded in connected and autonomous vehicles (CAVs) can handle the information generated by vehicular onboard devices, such as cameras, radars, light detection and ranging (LiDAR) sensors, ultrasonic sensors, global positioning systems (GPS), and so on \cite{OverviewITS}\cite{AVsurvey}. Through a series of deep learning-based computations, the driving scenarios—including path-planning, behavior arbitration, and motion control—can be realized.
Generally, deep learning requires massive data to first train a deep neural network (DNN) in a centralized manner so that the DNN learns to make decisions like humans \cite{Chenmingzhe}. After that, the DNN can be deployed or loaded into an application environment. Besides, the application environment also needs to be configured and updated in real time according to the global information of the driving environment. This information, which includes  weather conditions, navigations of other CAVs  \cite{Yangxiaolong},  driving conditions of road segments ahead, as well as high precision maps, can be regarded as fundamental data for the application. The practical application is mainly the process of inference (prediction) of the DNN, which inputs new data encountered in the real world into the pre-trained model, yielding results that can guide decision-making \cite{EIsurvey}. Due to the volume and variety of information, inference processes require relatively complicated computations.  In particular, for intelligent transportation systems (ITS) that have huge demands on resources, the design of computing schemes is more demanding and  challenging. 
%\footnote{The centralized training can operate at cloud centers or other powerful servers.}

Recently, multi-access edge computing (MEC) has been adopted in 5G networks \cite{Renqiqi}. Compared with centralized cloud computing, MEC provides computation capabilities in proximity, thus improving both delay and energy performance \cite{ZhouyuchenSurvey}. This technique makes it possible to push the heavy computational burden of AI to edge computing systems, which has given rise to the research field of edge intelligence (EI) \cite{EIsurvey}. Thus, assisting CAVs to perform AI tasks at edge servers has motivated the development of ITS to a certain extent \cite{Luoquyuan}. Autonomous driving will produce terabytes of data everyday\cite{Car}. Thus, the highly efficient processing and analysis–among other things in real-time–will be of great importance. Obviously, due to the limited  resources and coverage of edge nodes, traditional edge computing can no longer cope with this situation.
Much work has been done to find more suitable solutions, for instance, extending edge computing to aerial networks or more remote satellite networks \cite{compuUAV,UavTWC,SAGINTWC,JSACchengnan,SysjourSatellite}.
Unmanned aerial vehicle (UAV) computing was proposed to complement terrestrial networks, given that UAVs can be deployed flexibly at relatively low altitudes and offer line-of-sight communications. It is agreed that the UAV has limited energy, and therefore these works paid attention to minimizing energy consumption when computing tasks.  \cite{compuUAV} investigated to make offloading decisions, allocate transmitted bits in both uplink and downlink, as well as design UAV trajectory, and then block coordinate descent and successive convex approximation (SCA) methods were adopted to minimize the energy consumption.  Except offloading decision and trajectory planning of UAV, the work \cite{UavTWC} further investigated the CPU frequency control by adopting dynamic voltage and frequency scaling (DVFS) technique and utilized the SCA method to solve the energy minimization problem. However, in ITS, the UAV's trajectory planning needs to follow the mobility of CAVs, which will lead to additional energy consumption, and the limited computing power of the UAV cannot cope with computationally intensive vehicular tasks in a timely and efficient manner. To overcome the limitations of energy, coverage, and computing capability, more UAVs are needed to enhance coverage and computing, which obviously cannot be implemented in long-distance ITS.
	Recent cutting-edge publications have highlighted the potential of satellite networks for strong computational power and coverage and proposed space-air-ground integrated network (SAGIN). Based on this framework, the computing functionality of the ground network, aerial network (mostly UAV), and satellite network can be integrated to perform computing tasks. The authors in \cite{SAGINTWC} designed a task scheduling policy for IoT applications to minimize offloading and computing delay of all computing tasks, and it solved the problem by using DNN. Besides, a risk function is designed to ensure the energy consumption does not violate energy capacity constraint of UAV. The work \cite{JSACchengnan} proposed to leverage the framework of SAGIN to improve delay performance of remote IoT applications, where deep reinforcement learning method was used to solve task offloading of the whole network, and a heuristic algorithm was designed to solve task scheduling as well as computing resource allocation at the UAV. Although SAGIN computing has been verified to serve IoT applications effectively, it is not suitable for ITS applications. In addition to the prohibitive inherent high transmission delays between the satellite and the ground, mobility management will become a crucial issue when using SAGIN computing for ITS. Although the mobility management of UAV in SAGIN was considered by the work \cite{SysjourSatellite}, it is designed to serve static users, and the mobility of satellites was not discussed.  
	In fact, due to the mobility of satellites, UAVs, and CAVs, mobility management in such a system will be more complicated. In addition, the inherent high latency in satellite networks and the unreliability of connections in SAGIN are undoubtedly safety hazards for the delay- and stability-sensitive ITS applications. 

More recently, high altitude platform station (HAPS) systems have been proposed as candidates for 6G networks \cite{HAPSmagQiu,HanzhuJsac}.  The HAPS can be adopted as the leader of a stratospheric system operating at an altitude of around 20 km, providing line-of-sight communication and wide coverage with a radius of 50-500 km \cite{ITU}. It has a large payload (usually $\ge$ 100 kg), which allows being equipped with powerful computing resources and batteries. In addition, current and future energy conversion techniques for solar energy and wind power as well as battery techniques provide the HAPS with a powerful energy supply potential \cite{Gunes1,wind,battery}. The atmospheric temperature at the HAPS’ operating altitude is quite low, so it might not need too much energy for cooling \cite{Gunes1}. For these reasons, HAPS can
be regarded as a great complement to edge computing. The introduction of HAPS computing has three unique advantages for ITS. First, the HAPS can be flexibly deployed and integrated with terrestrial networks to fill the resource shortage, especially for congested and remote areas, without massive updating of terrestrial infrastructures\footnote{Here we utilize HAPS architecture to complement rather than replace terrestrial infrastructures.}.
Second, due to its wide coverage, long battery life as well as advanced energy conversion technologies, the HAPS can provide more stable and sustainable coverage for ITS, therefore reducing the frequent handoffs between CAVs and terrestrial infrastructures \cite{Sahabul}. 
Finally, the HAPS is suitable for collecting and storing fundamental data, such as the components requested for loading and configuring the ITS-based application environment. Note that since CAVs in the vehicular network may encounter the same problems and situations, the HAPS data library can apply to the entire intelligent transportation network under the coverage of the HAPS.

It follows from the above that computational performance of CAVs can benefit from the flexible access to the HAPS and RSUs in an integrated system, which means that the tasks of CAVs that need to be computed can be arranged at three destinations,  corresponding to three computational resources: onboard devices of CAVs, terrestrial network edge, and the HAPS. Then, as mentioned above, the HAPS can  collect and store fundamental data from a coverage area  in its data library. The required fundamental data should be obtained first to configure the computing environment at the computing destination, and then according to the individual input, a personalized AI inferencing (task computing) can be realized. Accordingly, in this work, we consider a ternary offloading strategy not partial offloading because partial offloading requires obtaining fundamental data at all destinations, which undoubtedly introduces additional delays.  When computing  at a CAV or a network edge is required, the fundamental data needs to be delivered from the HAPS to the computing destination. However, the long-distance transmission of data can cause large communication delays.  To address this issue, this work introduces the caching technique to store frequently requested fundamental data from the HAPS library at the network edge. Caching is a technique that can store data on nodes in advance, designed to quickly respond to requests and avoid backhaul or long-distance transmissions. Different from works \cite{cacheNOMA,cacheChenmingzhe}, this article considers the computation offloading and caching jointly.   The cached fundamental data can be accessed more quickly from the edge than from the HAPS, which consequently mitigates delays and improves performance.   Considering the dynamic nature of the network and the individual time-varying requirements, we need to reasonably determine the computation offloading policy for each CAV and plan the caching strategy of the fundamental data at network edges as well as to synergistically schedule computing and communication resources to improve delay performance for applications \cite{Weiyifei,3c,Gunes2}. The main contributions of this article are as follows:  
\begin{enumerate}
\item We introduce HAPS computing for ITS to improve delay performance. To the best of our knowledge, this is the first work that investigates the benefits of HAPS computing for ITS.% in accelerating task execution for ITS.
 \item By integrating the HAPS network with terrestrial edge networks, we develop a three-layer computation  framework, where CAVs can  leverage  the computing resources of their onboard devices, terrestrial network edges, and the HAPS to deal with the tasks at hand. In addition, by utilizing caching at network edges, we address the issue of large communication delays caused by  large-distance transmissions of fundamental data from HAPS.
\item With the objective of minimizing the total delay for executing ITS-based tasks, we formulate the optimization problem by considering the computation offloading policy for CAVs and caching strategy of fundamental data at network edges as well as the allocations of computing and bandwidth resources in the network. 
\item We decouple the formulated multi-slot mixed-integer and continuous variable problem into decision-making and resource allocation subproblems, where the former addresses computation offloading and caching decisions, and the latter addresses computing and bandwidth resource allocations. In addition, a multi-agent reinforcement learning method is used to solve the decision-making problem. Then, with the resulting decisions, we determine the optimal  resource allocation by using a Lagrangian method.
\end{enumerate}

The remainder of this article is organized as follows. In Section II, we present the system model and the optimization problem formulation.  Section III presents a method to solve the decoupled decision-making subproblem. Section IV provides the optimal solution of the decoupled resource allocation subproblem.  In Section V, we present the numerical results of the performance evaluation. Finally, Section VI offers some concluding discussions. 
\section{System Model And Optimization Formulation}
%In this section, we present the system model, including the proposed HAPS-assisted caching and computation offloading framework, computing model, caching model, communication model,  and delay model, followed by the optimization problem formulation. 
%\subsection{Proposed Framework}
	Fig. 1 shows the proposed HAPS-assisted caching and computation offloading framework for ITS. A single HAPS at an altitude of 20 km in the stratosphere provides coverage for several CAVs along a one-way highway\cite{TWC1}.  The road is divided into $M$ segments, labeled as $\mathcal{M}=\{1,2,\dots,M\}$, each of which is covered by an access point, i.e., roadside unit (RSU), and equipped with a server to compute tasks. The CAVs in the objective area are labeled as $\mathcal{I}=\{1,2,\dots,I\}$. In this article, we adopt the quasi-static assumption that the channel environment and network topology remain unchanged in one time slot. Due to the limited computational capabilities of their onboard devices, the CAVs experience serious delays in processing computing-intensive tasks by themselves. Thanks to the proximally situated RSUs and relatively resource-rich HAPS, we can consider a local-RSU-HAPS computation offloading framework for ITS\footnote{Hereafter, we call computing tasks at the onboard devices of CAVs and RSUs as local
		computing and RSU computing, respectively.}. In this framework, the fundamental data is stored in the HAPS library.  The individual data is generated by the onboard devices of CAVs. Note that the fundamental data is requested to configure the application environment quickly, and on this basis, personalized tasks can be computed when individual input data is ready. The sequence $\mathcal{T}=\{1,2,\dots,T\}$ represents the
times of successive decision slots. There are $N$ fundamental data, which are labeled as the library set $\mathcal{N}=\{1,2,\dots,N\}$. In time slot $t$, the volume of the fundamental data requested by CAV $i$ is represented by $\Omega_i(t)$,  the volume of individual data from CAV $i$ is denoted by ${{\varepsilon_i(t)}}$, and the volume of the computed result is denoted by $\phi_i(t)$. The location of CAV $i$ is indicated by $l_i(t)$.  Since the delay performance is critical for ITS, we focus on improving the total delay of all CAVs to complete the tasks. Therefore, the key factors, including the computation offloading decision as well as bandwidth and computing resource allocations, will be discussed. Besides, the caching capability of RSUs will also be explored so that the requested fundamental data can be pre-stored for future use to %further
 mitigate delays. A summary of the notation used in this article is summarized in Table I.
\begin{figure}[!t]
	%\centerline{\includegraphics[width=\columnwidth]{s.pdf}}
	\centerline{\includegraphics[width=10cm]{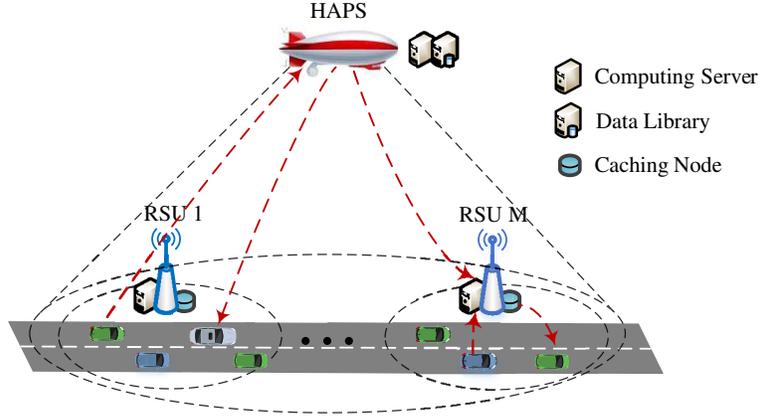}}
	\caption{HAPS-assisted caching and computation offloading framework for ITS.}
	\label{fig1}
	\vspace{-10pt}
\end{figure}
\begin{table}\nonumber
	\caption*{TABLE I: Notation Definitions}
	\vspace{-11pt}
	\begin{center}
		\begin{tabular}{c|c}
			\hline
			{Notation} & {Definition}  \\ 
				\hline
			{$\mathcal{M}, M$} & {The RSU set and the number of RSUs}  \\ 
				\hline
			{$\mathcal{I}, I$}& {The CAV set and the number of CAVs } \\ 
				\hline
			{$\mathcal{T} ,T$ }& {The decision time slot set and the number of decision time slots} \\ 
				\hline
			{$\mathcal{N} ,N$}& {The fundamental data set and the number of fundamental data}  \\ 
				\hline
			{$\Omega,\varepsilon,\phi$} & {Volume of fundamental data (Mbits), individual input data (Kbits), and computed result (Kbits)} \\ 
				\hline
			{$e$} &{ Computation density (CPU cycle/bit)}  \\ 
				\hline
			{$x,y,s$} & {Computation offloading decision, caching decision, and caching state of fundamental data} \\ 
				\hline		
			{$\pmb{\Pi}, \mathcal{J}$ } &{ Computing group sets and computing group index set} \\ 
				\hline
			{$F_{HAPS},F_{RSU},F_{CAV}$} &{ Computational capability of HAPS, RSU and CAV (CPU cycle/s)}  \\ 
			\hline
				{$G^{LoS}, G, G^{NLoS}$} & {LoS link channel gain, directional antenna gain, and NLoS link channel gain} \\ 
			\hline
				{$d^{LoS}, d^{NLoS}$} &{ LoS link distance and  NLoS link distance (m)} \\ 
			\hline
			{$c, f_c,\alpha, \beta_0$ }&	{Light speed (m/s), carrier frequency (Hz),  NLoS link path loss factor, and NLoS link reference path loss} \\ 
			\hline
				{$ h^{LoS}, h^{NLoS}$} &{LoS link small-scale fading coefficient  and NLoS link small-scale fading coefficient}  \\ 
			\hline
			{$\pmb{\Psi},\mathcal{K}$}&{ Communication group sets and  communication group index set}\\ 
			\hline
				{$N_0$} & {Gaussian noise power
				spectrum density (dBm/Hz)}  \\ 
			\hline
			{$B, R$}	 & {Bandwidth limitation (MHz) and transmission rate (bit/s) }\\ 
			\hline
			{$f, b$} &{ Computing resource allocation ratio and bandwidth allocation ratio} \\ 
			\hline
			{$P^T$ }& { Transmitter power (dBm) }  \\
			\hline
			{$T^L, T^R, T^H$} & {Delay under local, RSU, and HAPS computing (s) }  \\
		\hline
		\end{tabular}
	\end{center}
	\vspace{-4pt}
\end{table} 
\vspace{-10pt}
\subsection{Computing Model}
It is assumed that in each time slot $t$, each CAV requires only one personalized  task computing, which can be executed at only one destination. We use variable $x_i(t)$ to indicate the computation offloading decision of CAV $i$, where $x_i(t)=0,1,$ and $2$ corresponds to local, HAPS, and RSU computing, respectively.  Given offloading decision $\pmb{x}=\{x_i(t)|\forall i\in\mathcal{I}\}$, the CAVs are divided into $J$ groups according to their computing destinations, including one HAPS computing group, $M$ RSU computing groups, and one local computing group. The group sets are labeled as $\pmb{\Pi}=\{\Pi_j|j\in\mathcal{J}\}$, where $\mathcal{J}={\{HAPS\}\cup\{RSU_1,RSU_2,\dots,RSU_M\}\cup\{CAV\}}$ denotes the index set. As for the computing resources, we first assume that the computational capability for CAV is $F_{CAV}$ (in CPU cycle/s). Furthermore, the HAPS and each RSU are equipped with servers with computational capabilities of $F_{HAPS}$ and $F_{RSU_m|\forall m\in\mathcal{M}}$ (in CPU cycle/s), respectively. We consider that the computing resources are allocated at each RSU and HAPS according to the task workloads, which will be discussed in Section IV. Given this consideration, we let $f_{i,j}(t)$ indicate the allocated computing resource ratio of CAV $i$ in computing group $j$, and $\sum\limits_{i\in\Pi_j}f_{i,j}(t)\le 1, \forall j\in \{\mathcal{J}\setminus{CAV}\}$ limit the computing resource allocation at each server. Furthermore, the computation delay can be expressed as $\frac{\lambda_i(t)}{f_{i,j}(t)F_j}$ for $\forall j\in \{\mathcal{J}\setminus{CAV}\}$, and $\frac{\lambda_i(t)}{F_{CAV}}$ for $j=CAV$, where $\lambda_i(t)$ is the required workloads for processing CAV $i$'s task, obtained by the product of individual input data $\varepsilon_i(t)$ (in bit) and computation density  $e_i(t)$ (in CPU cycle/bit).  
\subsection{Caching Model}
Each RSU is equipped with storage, whose space is $C$ (in Mbits). The requested fundamental data, which is not in the storage, can be cached for future reuse. The storage follows the first-in-first-out policy, which means that when the content needs to be updated and the space is insufficient, the most previously stored content will be deleted \cite{Weiyifei}. We let binary variable $y_i(t)$ to indicate the decision of whether to cache the fundamental data requested by CAV $i$ at the end
of time slot $t$, where $y_i(t)=1$ means it will be cached at CAV $i$'s associated RSU, while $y_i(t)=0$ means it will not be cached\footnote{ Because of the correspondence of the fundamental data and the individual task, when a task is generated, the corresponding fundamental data will be requested to configure the application environment. }. Let binary variable $s_i(t)$ indicate the state of whether CAV $i$'s requested fundamental data is in the storage of its associated RSU, where $s_i(t)=1$ means it is in the storage, while $s_i(t)=0$ means it is not. When the task is performed by CAV $i$ or its associated RSU, it can receive a fast response if the requested fundamental data is  in storage; otherwise, it will get the content from the HAPS library.
\begin{figure}[!t]
	\centerline{\includegraphics[width=5cm]{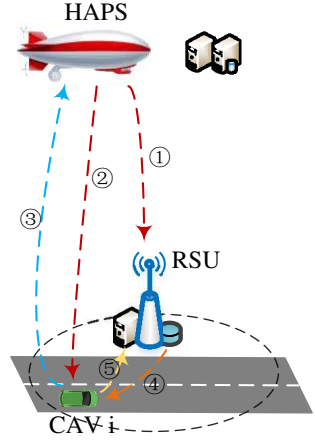}}
	\caption{Communication model.}
	\label{fig2}
\end{figure}  
\subsection{Communication Model}
 We consider both RSUs and CAVs are equipped with a single antenna, while the HAPS is equipped with multiple antennas, and the objective area represents one HAPS cell. Meanwhile, all links between the HAPS and CAVs, between the HAPS and RSUs, and  between RSUs and CAVs work on orthogonal channels\footnote{The channels are orthogonal in the immediate neighborhood, that is, the vicinity around the CAV is working in an orthogonal scheduling regime.}. Fig. 2 shows all links in the communication system.  Link \ding{172} is the downlink from the HAPS to the RSU for transmitting fundamental data; Link \ding{173} is the downlink from the HAPS to CAV $i$ for  transmitting fundamental data or the computed result; Link \ding{174} is the uplink from CAV $i$ to HAPS for offloading individual input data; Link \ding{175} is the downlink from the RSU to CAV $i$ for transmitting fundamental data or the computed result; Link \ding{176} is the uplink from CAV $i$ to the RSU for offloading individual input data. Since the HAPS  hovers in a high altitude and the highway is usually in a remote area, we consider line-of-sight (LoS) links for links \ding{172}, \ding{173} and \ding{174}, whose large-scale fading path loss follows the free space path loss, and the overall channel gain can be modeled by \cite{TWC2}:
 %For convenience, we utilize $L=ul,$ and $dl$ to distinguish uplink and downlink, and let $S=HV, HR,$ and $RV$ represent the links between HAPS and AV, between HAPS and RSU, and between RSU and AV (Here we consider the link type $S=RV$ means the link between the AV and its associated RSU), respectively.
\begin{equation}\small\label{op1}
G_i^{LoS}(t) = {G}{\left( {\frac{c}{{4\pi {d_i(t)}{f_c}}}} \right)^2}{\left| {{h_i^{LoS}(t)}} \right|^2}\quad i\in\mathcal{I}
\end{equation}
where  $c$ is the speed of light, $d_i(t)$ is the  distance of the link from the transmitter to the receiver.  $f_c$ is the carrier frequency, and  we consider 2 GHz for all links in this work.  $G$ is the directional antenna gain corresponding to the production of the transmitter antenna gain and receiver antenna gain. Since environment effects are negligible for the frequencies under 10 GHz, the environment attenuations are not considered in this work\cite{HAPsurvey2005}. $h_i^{LoS}(t)$ is the small-scale fading coefficient of the LoS links, corresponding to Rice fading.  

Links \ding{175} and \ding{176} are non-line-of-sight (NLoS) communication links, whose channel gains are modeled as follows:
\begin{equation}\small\label{op2}
G_i^{NLoS}(t)= \frac{{{\beta _0}({f_c}){{\left| {h_i^{NLoS}(t)} \right|}^2}}}{{{{\left( {{d_i}(t)} \right)}^\alpha }}}	
\end{equation}
 where $\beta_0({f_c})$ is the path loss at the reference distance 1 m,  $h_i^{NLoS}(t)$ is the small-scale  fading coefficient of the NLoS links following a Rayleigh distribution, and $\alpha$ is the path-loss exponent. 

To improve the delay performance, bandwidth resources are  allocated among links on the basis of channel gain and the volume of the transmitted data in this work, discussed in Section IV. To discuss the bandwidth allocation, we categorize all CAVs in the communication system into four groups, according to offloading decision $\pmb{x}$ and caching state $\pmb{s}$ of the requested fundamental data. The group sets of CAVs are labeled as $\pmb{\Psi}=\{\Psi_k|k\in\mathcal{K}\}$, where $\mathcal{K}=\{dl\_H,ul\_H,dl\_R,ul\_R\}$ is the index set of groups. By discussing the situation of CAV $i$, the group specifications and forming standards are listed as follows: 
\begin{enumerate}
\item ${dl\_H}$: The downlink transmission of the HAPS, including links \ding{172} and \ding{173}. For CAV $i$, link \ding{172} corresponds to $x_{i}(t)=2$, $s_{i}(t)=0$, where fundamental data needs to be transmitted from the HAPS to CAV $i$'s associated RSU.  Link \ding{173} corresponds to $x_i(t)=0, s_i(t)=0$ or $x_i(t)=1$, where the fundamental data or computed result will be transmitted from the HAPS to CAV $i$. The available bandwidth of CAVs in group $k=dl\_H$ is $B_{dl\_H}$.
\item ${ul\_H}$: The uplink transmission of the HAPS, corresponding to link \ding{174} with the offloading decision $x_i(t)=1$, where the individual input data will be transmitted from CAV $i$ to the HAPS. The available bandwidth of CAVs in group $k=ul\_H$ is $B_{ul\_H}$.
\item ${dl\_R}$: The downlink transmission of an RSU, corresponding to link \ding{175} with $x_i(t)=0, s_i(t)=1$ or $x_i(t)=2$, where the fundamental data or the computed results will be transmitted from the associated RSU to CAV $i$. The available bandwidth of CAVs in group $k=dl\_R$ is $B_{dl\_R}$.
\item ${ul\_R}$: The uplink transmission of the RSU, corresponding to link \ding{176} with $x_i(t)=2$, where the individual input data will be transmitted from CAV $i$ to its associated RSU.  The available bandwidth of CAVs in group $k=ul\_R$ is $B_{ul\_R}$.
\end{enumerate} 
Based on the above categorization, the channel gains can be restated as ${G_{i,k}}(t),\forall i\in\Psi_k, \forall k\in\mathcal{K}$, in which each composition is  from the channel gain given at either equation \eqref{op1} or equation \eqref{op2}.  
Therefore, the transmission rate can be given accordingly:
\begin{equation}\small\label{rate}
R_{i,k}(t) = b_{i,k}(t)B_k{\log _2}\left(1 + \frac{{{P^T_k}G_{i,k}(t)}}{{{b_{i,k}(t)}B_k{N_0}}}\right)
\end{equation}
where $b_{i,k}(t)$ is the ratio of bandwidth allocated to CAV $i$ in group $k$, and $\sum\limits_{i\in\Psi_k}b_{i,k}(t)\le 1,~ \forall k\in \mathcal{K}$ constrains the bandwidth ratios of the CAVs belonging to each group. $B_k$ is the available bandwidth of CAV $i$ in group $k$, and $N_0$ is the Gaussian noise power
spectrum density. With the transmission rate, the communication delay can be modeled as $\frac{D_{i}(t)}{R_{i,k}(t)}$ for CAV $i$ in group $k$, where $D_{i}(t)$ represents the volume of data to be transmitted, which includes fundamental data $\Omega_i(t)$, individual input data $\varepsilon_i(t)$, and the computed result $\varphi_i(t)$. 
\subsection{Delay Model}
Thus far, the computing, caching, and communication models have been formulated. Using these models, the delays of CAV $i$ under computing ways of local, HAPS, and RSU are given as follows:
\begin{enumerate}
\item Local computing:
\begin{equation}\small\label{op6}
\begin{split}
T_{i}^L(t)= {\mathbbm{1}_{{x_i(t)} = 0}} \cdot \left( {\frac{{{s_i(t)}{\Omega _i(t)}}}{{R_{i,dl\_R}(t)}} + \frac{{\big(1 - {s_i(t)}\big){\Omega _i(t)}}}{{R_{i,dl\_H}(t)}} + \frac{{{\lambda _i(t)}}}{{{F_{CAV}}}}} \right)
\end{split}
\end{equation}
\item HAPS computing:
\begin{equation}\small\label{op7}
\begin{split}
T_i^H(t)= {\mathbbm{1}_{{x_i(t)} = 1}} \cdot \left(\frac{{{\varepsilon _i(t)}}}{{R_{i,ul\_H}(t)}} + \frac{{\lambda _i(t)}}{{f_{i,HAPS}(t)F_{HAPS}}} + \frac{{{\varphi _i(t)}}}{{R_{i,dl\_H}(t)}}\right)
\end{split}
\end{equation}
\item RSU computing:
\begin{equation}\small\label{op6}
\begin{split}
T_i^R(t)= {\mathbbm{1}_{{x_i(t)} = 2}} \cdot \Bigg(\frac{{\big(1 - {s_i(t)}\big){\Omega _i(t)}}}{{R_{i,dl\_H}(t)}} + \frac{{{\varepsilon _i(t)}}}{{R_{i,ul\_R}(t)}}
+\frac{{\lambda {}_i(t)}}{{f_{i,RSU_m}(t)F_{RSU_m}|_{i\in{\Pi_{RSU_m}}}}}
+ \frac{{{\varphi _i(t)}}}{{R_{i,dl\_R}(t)}}\Bigg)
\end{split}
\end{equation}
\end{enumerate}
where $\mathbbm{1}_\Theta$ is a conditional indicator which equals $1$ when condition $\Theta$ is true; otherwise it equals $0$.  Note that there are no caching-related items in equation (5) because the requested fundamental data can be directly responded from the HAPS data library, regardless of the caching state at RSUs. 
\subsection{Optimization Problem Formulation}
Here, we formulate an optimization that optimizes the computation offloading decision $\pmb{x}$, caching decision $\pmb{y}$, bandwidth resource allocation $\pmb{b}$, and computing resource allocation $\pmb{f}$ for all CAVs, with the objective of minimizing the delays experienced by CAVs for executing tasks over multiple time slots. The optimization details are mathematically formulated as follows: 
\begin{subequations}\small\label{op8}
\begin{align}
 \min_{\substack{\pmb{x,y,b,f}}}&\ \sum\limits_{t\in\mathcal{T}}\sum\limits_{i\in\mathcal{I}} {T_i^L}(t)  + T_i^R(t) + T_i^H(t)\label{op8a}\\
&\;{x_i}(t) \in \{ 0,1,2\}\quad \forall i\in\mathcal{I},\ \forall t\in\mathcal{T}\label{op8b}\\
&\;{y_i}(t) \in \{ 0,1\}\quad \forall i\in\mathcal{I},\ \forall t\in\mathcal{T}\label{op8c}\\
&\sum\limits_{i\in\Psi_k} {b_{i,k}}(t) \le 1 \quad \forall k\in\mathcal{K},\ \forall t\in\mathcal{T}\label{op8d}\\
&\sum\limits_{i\in\Pi_j} {f_{i,j}(t) \le 1}\quad \forall j\in \mathcal{J},\ \forall t\in\mathcal{T}\label{op8e}\\
&b_{i,k}(t)> 0\quad \forall i \in {\Psi _k}, \ \forall k\in\mathcal{K},\ \forall t\in\mathcal{T}\label{op8f}\\
&f_{i,j}(t)> 0\quad \forall i \in {\Pi _j},\  \forall j\in\{\mathcal{J}\setminus{CAV}\},\ \forall t\in\mathcal{T}\label{op8g}
\end{align}
\end{subequations}
where \eqref{op8a} is the optimization objective function of the total delay of all CAVs over several time slots, which consists of the delays under three ways: local, RSU, and HAPS computing. \eqref{op8b} indicates the offloading decision of CAVs for executing tasks, and \eqref{op8c} indicates the caching decision of the requested fundamental data of CAVs. \eqref{op8d} denotes the bandwidth constraints for different types of links. \eqref{op8e} denotes the computing resource constraints of the HAPS and RSUs. \eqref{op8f} and \eqref{op8g} are positive requirements for resource allocation variables.  

Problem \eqref{op8} is a multi-slot mixed-integer and continuous variable problem with coupled computation offloading and caching decisions as well as bandwidth and computing variables, which is NP-hard to solve. In the following, we decouple the optimization problem into two subproblems, where we first adopt the multi-agent reinforcement learning method to solve the decision-making problem over several time slots. Then, we find the efficient solutions of the bandwidth and computing resource allocations in each time slot with the given decisions. 

\section{The MARL-based Decision-Making Problem for Computation Offloading and Caching}
We first discuss the decision-making problem for computation offloading and caching decisions by adopting the assumption of equal communication and computing resource allocations.  The decision-making optimization problem is given as follows:
\begin{subequations}\small\label{op9}
\begin{align}
\min_{\substack{\pmb{x,y}}}&\ \sum\limits_{t\in\mathcal{T}}\sum\limits_{i\in\mathcal{I}} {T_i^L}(t)  + T_i^R(t) + T_i^H(t)\label{op9a}\\
&\;{x_i}(t) \in \{ 0,1,2\}\quad \forall i\in\mathcal{I},\ \forall t\in\mathcal{T}\label{op9b}\\
&\;{y_i}(t) \in \{ 0,1\}\quad \forall i\in\mathcal{I},\ \forall t\in\mathcal{T}.\label{op9c}
\end{align}
\end{subequations}
This allows us to determine the computation offloading and caching for maximizing the long-term rewards. We will use a multi-agent deep reinforcement (MARL) learning method to solve this problem, where each CAV represents an agent making the offloading and caching decisions. 
Multi-agent systems (MAS) have been used in a wide range of fields, including telecommunications, distributed control, resource management, and collaborative decision support systems \cite{MAS}.  Reinforcement learning (RL) can help solve multi-agent problems over multiple time slots by enabling agents to interact sequentially with the environment, take actions, and perceive states and rewards to maximize cumulative rewards \cite{RLsurvey}. Depending on the relationship between the agents, MARL can be divided into three types: cooperative, competitive, and mixed. In this article, we consider the proposed architecture as a cooperative MARL problem where agents are independent decision-makers who coordinate their actions and work together to achieve a shared team reward accumulated over time \cite{MAS}. 
\subsection{Dec-POMDP}
The training process of this MARL is modeled as a decentralized partially observable Markov decision process (Dec-POMDP), whose main components are defined as follows:
\begin{enumerate}
\item Actions: At each time slot $t$, each agent
$i\in \mathcal{I}$ chooses an action $u_i(t)\in \mathcal{U}$, forming
a joint action $\pmb{u_t}\in \pmb{\mathcal{U}}$. 
In this article, each agent $i$ needs to make offloading and caching decisions, and therefore the action space of $\mathcal{U}$ derives from combinations of $x_i(t)=0,1,$ or $2$ and $y_i(t)=0$ or $1$. Note that when agent $i$ chooses either local computing or HAPS computing, we do not need to consider whether to cache the requested fundamental data at the RSU; therefore, the action space $\mathcal{U}$ can be defined as four elements for each agent. Here, we define $u_i(t)=0$, and $1$ to represent agent $i$  choosing local computing and HAPS computing for its task, respectively. And we define $u_i(t)=2$ for the case that the agent $i$ chooses RSU computing for its task and the requested fundamental data will not be cached at the associated RSU at the end of the time slot $t$. $u_i(t)=3$ means that agent $i$ will choose RSU computing and the requested fundamental data will be cached. 
\item Observations: $o_{i}(t)$ is the observation of agent $i$, and $\pmb{o_t}$ is a joint observation. Here we regard the joint observation $\pmb{o_t}$ as the global state of the whole system.  During execution, each agent perceives its own observation independently and makes the independent decision without additional communication. The observation space includes: (a) The location information $l_i(t)$ \cite{Luoquyuan}; (b) The index of the requested fundamental data $n_i(t)$, and the flag $s_i(t)$ indicating whether the requested content of agent $i$ is stored with its associated RSU; (c) The individual input data volume $\varepsilon_i(t)$, the output data volume $\varphi_i(t)$, and the computation density $e_i(t)$ \cite{Weiyifei} \footnote{Since given the present observation, the future observation is independent of the past observation, the proposed system conforms to Markovian characteristics. }.  
\item Team reward: In this system, each agent receives the joint team reward despite their individual observations and separate actions. Considering the objective is to reduce the total delay of CAVs, we use the sigmoid function to normalize the negative value of the total delay as the reward value ${r}_t$. 
\end{enumerate}   

With these definitions, the sequential decision-making optimization problem in a dynamic environment can be formulated as a Dec-POMDP. %The goal of all agents is to maximize the expected cumulative discounted joint reward, i.e., ${\mathcal{R}_t} = \sum\nolimits_{m = 0}^\infty  {{\gamma ^m}{r_{t + m}}}$, where $\gamma$ is the discount factor.
\subsection{Solving the Problem with Value-Decomposition Networks }
One commonly used method in MAS is the independent Q-learning (IQL) algorithm, which decomposes the multi-agent problem into a set of simultaneous single-agent problems in a shared environment with multiple agents \cite{VDN2}. In this case, each agent learns by taking the other agents as part of the environment. In practice, however, other agents in the environment are not fixed and constantly learn from their recent experience. Besides,  the partial observability of the agents prevents them from observing the global state of the environment.  For example, under the same observation $o_i(t)$, although agent $i$ can take the same action, the joint team reward obtained may be different because the global state may be different, and the policies of other agents are always changing, which leads to non-stationarity across the whole system. To address this issue, we adopt a centralized training decentralized execution (CTDE) framework in the collaborative MAS, which allows the global state to be acquired during the centralized training, and the agents can make independent decisions based on local observations only. During training, the optimal decision policy is trained globally. Given a policy $\pi$, the joint action-value function (Q function) is defined as the expected cumulative rewards of the global state $\pmb{o}$ and the joint action $\pmb{u}$. i.e., $Q_{tot}^\pi (\pmb{o},\pmb{u}) =  \mathbb{E}\left[ {\sum\limits_{m = 0}^\infty  {{\gamma ^m}{r_{t + m}}\left| {{\pmb{o_t}} = \pmb{o},{\pmb{u_t}} = \pmb{u};\pi } \right.} } \right]$, where $\gamma$ is the discount factor \cite{VDN2}.  However, a joint action-value function $Q_{tot}(\pmb{o},\pmb{u})$ obtained during the centralized training will not be exploited during distributed execution as only local observations can be acquired for agents.  A new proposal of value decomposition networks (VDNs) can be applied during the training process to decompose $Q_{tot}(\pmb{o},\pmb{u})$ into individual action-value functions.  VDN is an enhancement that builds on the deep Q network (DQN) to overcome the issue mentioned above\cite{VDN2}. In partially observed environments, a policy for the agent $i$ maps from histories to actions because the histories contain all information an agent has \cite{Book,VDN2}. Therefore, the action-value function of agent $i$ is ${Q_i}({\tau_i},{u_i}), \forall i\in \mathcal{I}$, where $\tau_i$  is the action-observation history (replacing global state) defined as ${\tau_i(t)}=u_i(1)o_i(1)r_1,...,u_i(t-1)o_i(t-1)r_{t-1}$ and $u_i$ is the action. With VDN, the joint action-value function is approximately the sum of all independent action-value functions  which is only relevant to individual action-observation history, i.e., 
\begin{equation}\small\label{vdn}
Q_{tot}^{\pi}(\pmb{o},\pmb{u}) \approx \sum\limits_i^I {{Q_i^{\pi}}({\tau_i},{u_i};\theta_i)} 
\end{equation} 
where $\theta$ is the parameter of the network as in DQN. Building on DQN, the VDN algorithm adopts the fixed target network and an experience replay buffer, and it utilizes the joint reward to iteratively update network parameters to minimize the loss function: $L(\theta) = \mathbb{E}\left[ {{{\left( {{y^{tot}} - {Q_{tot}}({\pmb{o _t}},\pmb{{u_t}})} \right)}^2}} \right].$
This expectation is over the mini-batches drawn from the experience replay buffer. The target value is defined by ${y_t^{tot}} = {r_t} + \gamma \mathop {\max }\limits_{{u_{t + 1}}} {Q_{tot}}(\pmb{{o_{t + 1}}},\pmb{{u_{t + 1}}})$. 
The gradients are backpropagated following the Q-learning rule to update ${Q_i}$ for each agent $i$. As a result, the VDN algorithm allows the use of global information during the centralized training and enables agents to execute actions independently according to local observations. In other words, during the  execution, the trained model can be applied to CAVs to make decisions in each time slot according to individual observations.
\section{Optimal Allocation for Bandwidth and  Computing Resources}
In this section, we investigate the optimal bandwidth and computing resource allocation problem in \eqref{op8} with the given computation offloading policy and caching strategy. Concretely, we utilize the Lagrangian dual method to formulate the Karush-Kuhn-Tucker (KKT) conditions of the problem, and then we find the closed-form solution for bandwidth resource allocation, and the optimal solution for computing resource allocation by using a bisection search method. 

Since the resource allocation problem is developed within one time slot, for brevity, we omit ``$(t)$" below, e.g., $f_{i,j}$ will represent $f_{i,j}(t)$, unless the time slot $t$ is emphasized.
Assuming the computation offloading and caching decisions are given, we first rewrite the resource allocation problem as follows:
\begin{subequations}\small\label{op10}
\begin{align}
\min_{\substack{\pmb{b,f}}}&\ \sum\limits_{k \in\mathcal{K}} {\sum\limits_{i \in {\Psi _k}} {\frac{{{O_{i,k}}}}{{{b_{i,k}}{{\log }_2}(1 + \frac{{{H_{i,k}}}}{{{b_{i,k}}}})}}} }  + \sum\limits_{j \in\{\mathcal{J}\setminus{CAV}\}} {\sum\limits_{i \in {\Pi _j}} {\frac{{{U_{i,j}}}}{{{f_{i,j}}}}} }+ \sum\limits_{i \in \Pi_{CAV} }  {\frac{{{\lambda _i}}}{{{F_{CAV}}}}} \label{op10a}\\
&\sum\limits_{i \in {\Psi _k}} {{b_{i,k}} \le 1}\quad  \forall k\in\mathcal{K} \label{op10b}\\
&\sum\limits_{i \in {\Pi _j}}{{f_{i,j}}}  \le 1\quad  \forall j\in\{\mathcal{J}\setminus{CAV}\}\label{op10c}\\
&b_{i,k}> 0\quad \forall i \in {\Psi _k}, \ \forall k\in\mathcal{K}\label{op10d}\\
&f_{i,j}> 0\quad \forall i \in {\Pi _j},\  \forall j\in\{\mathcal{J}\setminus{CAV}\}\label{op10e}.
\vspace{-4pt}
\end{align}
\end{subequations}
In \eqref{op10a}, the first term denotes the communication delays, where $O_{i,k}=\frac{D_{i}}{B_{k}}$ and $H_{i,k}=\frac{P_k^TG_{i,k}}{B_{k}{N_0}}$. The second term denotes the computation delays at the servers, where $U_{i,j}=\frac{\lambda_i}{F_{j}}$. The third term denotes the computation delays at the CAVs, which can be omitted in the further analysis of the problem \eqref{op10} since the available computational capabilities of the CAVs are fixed in this work. \eqref{op10b} to \eqref{op10e} are constraints of bandwidth and computing resource allocations.  \\
\textbf{Proposition 1:} Problem \eqref{op10} is convex. 
\begin{proof}
For simplicity, we first define functions $w(b_{i,k})=b_{i,k}\log_2\left(1+\frac{H_{i,k}}{b_{i,k}}\right)$, $g(b_{i,k})=\frac{O_{i,k}}{w(b_{i,k})}$, and $z(f_{i,j})=\frac{U_{i,j}}{f_{i,j}}$. As a consequence, the second-order derivatives can be calculated as 
\begin{equation}\small\label{eq1}
w''({b_{i,k}}) =  - \frac{{H_{i,j}^2}}{{\ln 2 \times {b_{i,k}}{{({b_{i,k}} + {H_{i,k}})}^2}}}
\end{equation}
\begin{equation}\small\label{eq2}
g''({b_{i,k}}) = \frac{{{O_{i,k}}\left[ {2{{(w'({b_{i,k}}))}^2} - w''({b_{i,k}})w({b_{i,k}})} \right]}}{{{w^3}({b_{i,k}})}}
\end{equation}
\begin{equation}\small\label{eqz}
z''({f_{i,j}}) = \frac{{2{U_{i,j}}}}{{f_{i,j}^3}}.
\end{equation}
From \eqref{eq1}, we can observe that $w''(b_{i,k})<0$, and therefore $g''({b_{i,k}})>0$ holds and the  convexity  of $g({b_{i,k}})$  can be obtained. From \eqref{eqz}, we can obtain the convexity of $z(f_{i,j})$ and the proof is completed. 
\end{proof}
Since problem \eqref{op10} is a standard convex optimization problem, we can solve it by using general convex optimization methods such as the interior point method. However, the complexity of the interior point method increases quickly with the number of CAVs. Now we will use Lagrangian dual method to find the solution efficiently.  Let us introduce the Lagrangian multipliers for constraints \eqref{op10b} and \eqref{op10c} to formulate a partial Lagrangian function:
\begin{equation}\small\label{Lagrangian}
\begin{split}
\mathcal{L}({{\pmb{b},\pmb{f}}},\pmb{\eta},\pmb{\mu}) 
=\sum\limits_{k \in\mathcal{K}} {\sum\limits_{i \in {\Psi _k}} {g({b_{i,k}})} }+\sum\limits_{j \in\{\mathcal{J}\setminus{CAV}\}} {\sum\limits_{i \in {\Pi _j}}z(f_{i,j}) } 
+\sum\limits_{k \in\mathcal{K}}\eta_k(\sum\limits_{i \in {\Psi _k}} {{b_{i,k}}}  - 1)  + \sum\limits_{j \in\{\mathcal{J}\setminus{CAV}\}}\mu_j(\sum\limits_{i \in {\Pi _j}} {{f_{i,j}}}  - 1). 
\end{split}
\end{equation}
The Lagrangian dual function is given by $\mathcal{D}({{\pmb{\eta},\pmb{\mu}}}) 
=\min_{\substack{\pmb{b,f}}}\{\mathcal{L}(\pmb{b},\pmb{f},\pmb{\eta },\pmb{\mu})\ |\ \pmb{b> 0},\pmb{f> 0}\}$,
and the dual problem is $\max_{\substack{\pmb{\eta,\mu}}}\ \{\mathcal{D}(\pmb{\eta},\pmb{\mu})\ |\ \pmb{\eta\ge 0},\pmb{\mu\ge 0}\}.$ 
As \eqref{op10} is a convex problem, the dual problem can achieve the same optimal objective with the primal problem by the strong duality \cite{cvx}. In the following, we provide the analysis of the KKT conditions of \eqref{op10}, which are sufficient and necessary for the optimal solution and yield a computationally efficient algorithm. The KKT conditions are expressed as follows:
\begin{subequations}\small\label{op11}
\begin{align}
&\nabla_{b_{i,k}}\mathcal{L}(\pmb{b,f,\eta,\mu})=0\quad \forall i\in\Psi _k,\ \forall k\in\mathcal{K}\label{op11a}\\
&\nabla_{f_{i,j}}\mathcal{L}(\pmb{b,f,\eta,\mu})=0\quad \forall i\in\Pi_j,\  \forall {j \in\{\mathcal{J}\setminus{CAV}\}}\label{op11b}\\
&\eta_k(\sum\limits_{i \in {\Psi _k}} {{b_{i,k}}}-1)=0\quad \forall k\in\mathcal{K}\label{op11c}\\
&\mu_j(\sum\limits_{i \in {\Pi _j}} {{f_{i,j}}} -1)=0\quad \forall {j \in\{\mathcal{J}\setminus{CAV}\}}\label{op11d}\\
&\sum\limits_{i \in {\Psi _k}} {{b_{i,k}}}  - 1\le 0\quad \forall k\in\mathcal{K}\label{op11e}\\
&\sum\limits_{i \in {\Pi _j}} {{f_{i,j}}}  - 1\le 0 \quad \forall {j \in\{\mathcal{J}\setminus{CAV}\}}\label{op11f}\\
&b_{i,k}> 0\quad \forall i \in {\Psi _k}, \ \forall k\in\mathcal{K}\label{op11g}\\
&f_{i,j}> 0\quad \forall i \in {\Pi _j},\  \forall j\in\{\mathcal{J}\setminus{CAV}\}\label{op11h}\\
&\eta_k\ge 0\quad \forall k\in\mathcal{K}\label{op11i}\\
&\mu_j\ge 0 \quad \forall {j \in\{\mathcal{J}\setminus{CAV}\}}.\label{op11j}
\end{align}
\end{subequations}
Here, \eqref{op11a} and \eqref{op11b} are necessary conditions for the feasible solution of the Lagrangian function; \eqref{op11c} and \eqref{op11d} are relaxation complementary conditions; \eqref{op11e} to \eqref{op11h} are primal constraints; \eqref{op11i} and \eqref{op11j} are the conditions that should be met for the Lagrangian multipliers of inequality constraints. 

We use the definitions of  $g(b_{i,k})={ {\frac{{{O_{i,k}}}}{{{b_{i,k}}{{\log }_2}(1 + \frac{{{H_{i,k}}}}{{{b_{i,k}}}})}}} }$ and $z(f_{i,j})=\frac{U_{i,j}}{f_{i,j}}$ to make further discussions. According to \eqref{op11a} and \eqref{Lagrangian}, we can obtain that 
\begin{equation}\small\label{root}
\begin{split}
g'(b_{i,k})+\eta_k=0\quad \forall i\in\Psi _k,\  \forall k\in\mathcal{K}.
\end{split}
\end{equation} 
\textbf{Proposition 2:} The root $\eta_k^*$ of \eqref{root} is unique and positive. 
\begin{proof}
 Recalling \eqref{eq2} in the Proof of Proposition 1, $g''({b_{i,k}})>0$, which indicates the first-order derivative of $g(b_{i,k})$ is a monotonically increasing function, calculated as follows: 
\begin{equation}\small\label{fistg}
\begin{split}
g'(b_{i,k})= -\frac{{\ln 2 \times {O_{i,j}}\left( \left( {{b_{i,j}} + {H_{i,j}}} \right){\ln \left( {1 + \frac{{{H_{i,j}}}}{{{b_{i,j}}}}} \right) - {H_{i,j}}} \right)}}{{b_{i,j}^2\left( {{b_{i,j}} + {H_{i,j}}} \right){{\ln }^2}\left( {1 + \frac{{{H_{i,j}}}}{{{b_{i,j}}}}} \right)}}.
\end{split}
\end{equation}
Therefore, we can obtain $g'(b_{i,k})\le g'(1)$. To determine $g'(b_{i,k})$, we need to discuss $g'(1)$, which is expressed by 
 $g'(1)= - \frac{{\ln 2 \times {O_{i,j}}\left( {\left( {1 + {H_{i,j}}} \right)\ln \left( {1 + {H_{i,j}}} \right) - {H_{i,j}}} \right)}}{{\left( {1 + {H_{i,j}}} \right){{\ln }^2}\left( {1 + {H_{i,j}}} \right)}}$.
It is difficult to observe whether $g'(1)$ is a negative or a positive value directly. Now, we further let $\mathcal{Z}(H_{i,j})={\left( {1 + {H_{i,j}}} \right)\ln \left( {1 + {H_{i,j}}} \right) - {H_{i,j}}}$. The first-order derivative of $\mathcal{Z}(H_{i,k})$ is calculated as $\mathcal{Z}'(H_{i,k})=\ln(1+H_{i,k})$.
It is clear that $\mathcal{Z}'(H_{i,k})>0$, which indicates function $\mathcal{Z}(H_{i,k})$ monotonously increases with $H_{i,k}$. Consequently, $\mathcal{Z}(H_{i,k})>\mathcal{Z}(0)=0$ (Note that $H_{i,k}>0$). After that, we can obtain $g'(1)<0$. And accordingly $g'(b_{i,k})<0$ holds, where $b_{i,k} \in (0,1]$. From \eqref{root}, we can obtain $\eta_k>0$ can be always met. By combining the monotonically increasing property of $g'(b_{i,k})$, we can conclude that the root $\eta_k^*$ of \eqref{root} is always unique and positive.  
\end{proof}
With Proposition 2, a bisection search method can be used to obtain the optimal $\eta_k^*$ \cite{Bisuzhi,solution}. 
Given that $\eta_k>0$ always holds, we know that the equality $\sum\limits_{i\in\Psi_k}{{b_{i,k}}}-1=0$ for $\forall k\in\mathcal{K}$ must hold. Then, the bandwidth allocation solution can be obtained by solving equation \eqref{root} iteratively  over $\eta_k\in(0,\eta_{\max}]$. The detailed process of the bisection search method is outlined in Algorithm 1. Given a precision $\delta$, Algorithm 1 requires $O(\log_2(\frac{\eta_{\max}}{\delta}))$ number of iterations to converge. 
\begin{algorithm}
\caption{Bandwidth Allocation Solution: Bisection Search Method}
\label{alg:A}
\begin{algorithmic}[1]
\STATE For each group $k\in\mathcal{K}$, initialize ${\eta_{k,l}=0}$, $\eta_{k,h}=\eta_{\max}$, $\delta>0 $, and set $q=0$.
\WHILE {$\left| {{\eta _{k,h}}-{\eta _{k,l}}}\right| \ge \delta $}
\STATE ${\eta _{k,q}} = (\eta_{k,h}  + {\eta _{k,l}})/2$ and $q=q+1$.
\STATE Obtain the root $b_{i,k}$ of the equation $g'(b_{i,k})+\eta_{k,q}=0$ for each $i\in\Psi_k$ by using a bisection numerical method. 
\IF {$\sum\limits_{i\in{\Psi_k}} {{b_{i,k}} > 1}$} 
\STATE let $\eta_{k,l}=\eta_{k,q}$.
\ELSE 
\STATE let $\eta_{k,h}=\eta_{k,q}$.
\ENDIF
\ENDWHILE
\end{algorithmic}
\end{algorithm} \par
Next, we will discuss how to solve the computing resource allocation. Observing the KKT conditions in \eqref{op11}, we know constraints \eqref{op11b}, \eqref{op11d}, \eqref{op11f}, \eqref{op11h}, and \eqref{op11j} are related to computing resource allocation variable $\pmb{f}$. To make it clear, in the following, we restate these constraints as
\begin{subequations}\small\label{op12}
\begin{align}
 &- \frac{{{U_{i,j}}}}{{{f_{i,j}}^2}} + {\mu _j} = 0\quad \forall i \in \Pi_j,\  \forall{j \in\{\mathcal{J}\setminus{CAV}\}} \label{op12a}\\
&\mu_j(\sum\limits_{i \in {\Pi _j}} {{f_{i,j}}}  - 1)=0 \quad \forall{j \in\{\mathcal{J}\setminus{CAV}\}}\label{op12b}\\
&\sum\limits_{i \in {\Pi _j}} {{f_{i,j}}}  - 1\le 0\quad \forall{j \in\{\mathcal{J}\setminus{CAV}\}}\label{op12c}\\
&f_{i,j}>0\quad \forall i \in \Pi_j,\ \forall{j \in\{\mathcal{J}\setminus{CAV}\}}\\
&\mu_j\ge 0 \quad \forall {j \in\{\mathcal{J}\setminus{CAV}\}}.
\end{align}
\end{subequations}
From \eqref{op12a}, we can conclude that $\frac{{{U_{i,j}}}}{{{f_{i,j}}^2}}  = {\mu _j}, ~\forall i\in {\Pi _j},\ \forall{j \in\{\mathcal{J}\setminus{CAV}\}}.$
Accordingly, we know that $\mu_j>0$ always holds (Note that both ${{U_{{i},j}}}$ and ${{f_{{i},j}}} >0$ for $\forall{i}\in {\Pi _j}$). As a result, the following expression must hold: 
\begin{equation}\small\label{eq4}
\sum\limits_{i \in {\Pi _j}} {{f_{i,j}}}  - 1=0 \quad \forall{j \in\{\mathcal{J}\setminus{CAV}\}}.
\end{equation}  
From $\frac{{{U_{i,j}}}}{{{f_{i,j}}^2}}  = {\mu _j}, ~\forall i\in {\Pi _j},\ \forall{j \in\{\mathcal{J}\setminus{CAV}\}}$, we can obtain the following formula:
\begin{equation}\small\label{eq5}
\begin{split}
&{f_{i,j}} = \sqrt {\frac{{{U_{i,j}}}}{{{\mu _j}}}} \quad \forall i\in {\Pi _j}, \ \forall{j \in\{\mathcal{J}\setminus{CAV}\}}.
\end{split}
\end{equation}
Substituting \eqref{eq5} into \eqref{eq4}, we can know that
\begin{equation}\small\label{eq6}
\frac{1}{{\sqrt {{\mu _j}} }}\left( {\sum\limits_{i \in {\Pi _j}} {\sqrt {{U_{i,j}}} } } \right) = 1\quad \forall{j \in\{\mathcal{J}\setminus{CAV}\}}.
\end{equation}
Therefore, we can obtain the closed-form optimal solution for the computing resource allocation as follows:
\begin{equation}\small\label{eq7}
{f_{i,j}} = \frac{{\sqrt {{U_{i,j}}} }}{{\sum\limits_{i \in {\Pi _j}} {\sqrt {{U_{i,j}}} } }}\quad \forall i \in \Pi _j,\ \forall{j \in\{\mathcal{J}\setminus{CAV}\}}.
\end{equation}
%, which achieves lower complexity

 Up to now, the specific solving process of problem \eqref{op8} in each time slot is described in two steps. The first step is to apply the trained VDN network model to each CAV so that it can select a computing destination for its task based on its own observations. If RSU computing is selected, it will also determine whether to cache the corresponding fundamental data in its associated RSU at the end of the time slot for future use. In the second step, under the offloading decision determined in the first step and the current caching state of fundamental data at RSU, the Lagrangian method will be used to solve the resource allocation sub-problem of the current time slot.

 In this work, we proposed a new paradigm for ITS which considers the average delay performance for the whole network. Moreover, our work can be easily extended for prioritizing the delay performance of specific CAVs, by changing the objective function into the weighted summation of the delay of all CAVs, and adding a constraint to guarantee a maximum delay for each CAV. The proposed solution in this work can still solve this problem by easily setting a penalty term in the reward to capature whether the new constraint is violated.
\section{Simulation Results}

In this section, we present the simulation results to evaluate the performance of our proposed scheme. 
We consider a 400 m one-way road covered by two RSUs and one HAPS. The HAPS hovered in the air at an altitude of 20 km and over the middle of the road horizontally. Two RSUs were deployed at the positions of 100 m and 300 m, each of which covers a range of 200 m. In this system, ten CAVs were initially randomly deployed on the road and their speed was set as 10 m/s. There are $N=200$ fundamental data in the HAPS library, whose data amounts were set as 2 Mbits, 5 Mbits, and 8 Mbits randomly in one-time simulation realization. We assumed that the contents in the library were sorted on the basis of their popularity. In each time slot, the requesting content popularity distribution was modeled as a Zipf distribution.% and the shape factor $\rho$ was 0.1.  
Each CAV randomly generated one task to be computed which depends on the individual input data per time slot and requested the corresponding fundamental data. If not emphasized, the main parameters of tasks were randomly set as follows. The individual data amount was set as 200 Kbits, 500 Kbits, and 800 Kbits, the computation density was set as 500, 1000, and 1,500 in cycle/bit, and the output data amount was set as 60 Kbits and 90 Kbits. The computational capabilities of the CAV, RSU, and HAPS were set as 2 G, 16 G, and 50 G in CPU cycle/s, respectively. The caching space of RSU was set as 600 Mbits. For communication parameters, the transmitting powers of CAV, RSU, and HAPS were set as 23 dBm, 27 dBm, and 33 dBm \cite{33dbm}, respectively, and the available bandwidths $B_{ul,R}$, $B_{ul,H}$, $B_{dl,R}$, and $B_{dl,H}$ were set as 5 MHz, 5 MHz, 20 MHz, and 20 MHz, respectively. The noise power spectral density was -174 dBm/Hz, and the directional antenna gain for HAPS was 17 dBi\cite{17dbi}. For LoS links, the small-scale fading followed a Rician distribution with rician factor 10 dB, and for NLoS links, the path-loss  exponent $\alpha$ was 3.7, and the small-scale fading followed a Rayleigh distribution $\mathcal{CN}(0,1)$.  

\begin{figure}[!t]
	%\centerline{\includegraphics[width=\columnwidth]{s.pdf}}
	\centerline{\includegraphics[width=7cm]{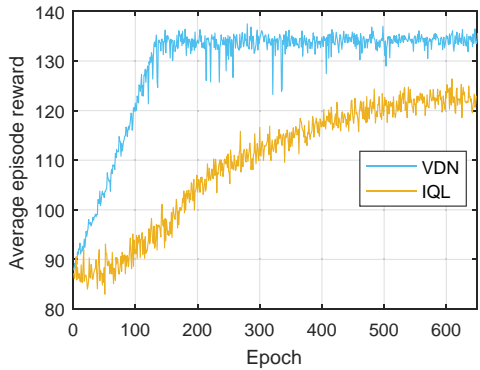}}
	\caption{Convergence of VDN algorithm.}
	\label{fig3}
\end{figure}

Here we show the simulation settings of the multi-agent reinforcement learning and bisection search methods. The learning process of the agents in both IQL and VDN algorithms was on the basis of the DQN, including the typical techniques of experience replay, target networks, and back propagations. The agent architecture followed \cite{VDN2}, where the inputs were processed first by a fully connected linear layer followed by a nonlinear activation function ReLU and then a GRU-cell, which is a kind of sequential RNN layer, and finally a linear layer was adopted to output the individual $Q$ values. The agents in the IQL algorithm trained their own networks in a decentralized manner, while following the CTDE mode and principle in \eqref{vdn}, the VDN algorithm utilized the summation of individual $Q$ values for centralized training to optimize the team reward. The number of units in hidden layers was set to 128. We set the learning rate and discount factor as 0.0005 and 0.95, and the minibatch and replay buffer sizes as 64 and 10,000. Moreover, the episode limitation $T$ was 150, and each epoch consisted of eight episodes. As for the bandwidth allocation algorithm, the  initializing multiplier $\eta_{\max}$ and the convergence precision $\delta $ were set as 50 and $10^{-8}$, respectively.         

\begin{figure}[!t]
	%\centerline{\includegraphics[width=\columnwidth]{s.pdf}}
	\centerline{\includegraphics[width=7cm]{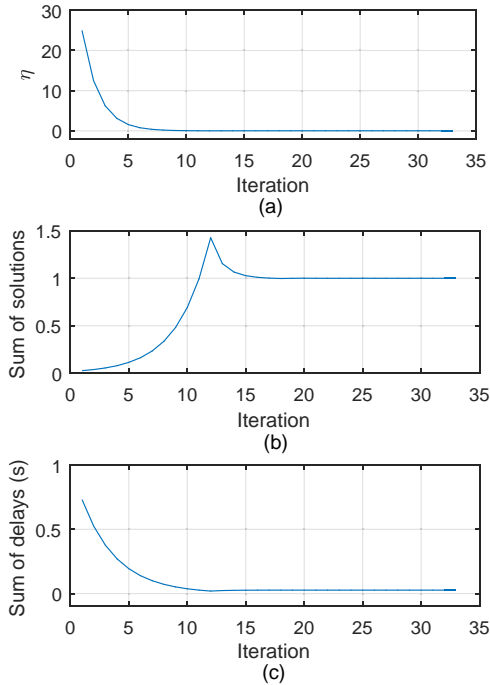}}
	\caption{Convergence and problem-solving process of Algorithm 1.}
	\label{fig4}
\end{figure}

Fig. 3 shows the convergence performance of the VDN algorithm with the IQL algorithm as the benchmark. The evaluation is based on the average episode reward of eight episodes in each epoch. The training process of the IQL algorithm is unstable and slow, and it is difficult to learn a good policy. This is because all the agents train independently and can only observe the environment locally and make decisions from their local perspectives. However, the environment changes dynamically with the policy and observation of each agent. Even if each agent takes the same action based on the same local observation, the environment changes dramatically due to the various observations and actions of other agents; so it is unstable and difficult to learn, and converges to a lower reward than the VDN algorithm. Because of its centralized training and decentralized execution (CTDE), the VDN algorithm can efficiently update the policy for each agent from a global point of view, so that each agent can learn to take action by taking into account the global situation even with local observations. Therefore, both the efficiency and stability are enhanced. 
%Fortunately, the proposed scheme has the advantage of addressing real-time network states and making decisions considering future benefits.

Fig. 4 shows the convergence performance and problem-solving process of Algorithm 1. Specifically, by Fig. 4(a), we  know the multiplier $\eta$ decreases with the iterations  initially, and it reaches convergence in about the 20-th iteration.  Fig. 4(b) shows the details of the summation of solutions\footnote{Here the solution means the result of bandwidth allocation ratio $b_{i,k}$. For each group $k$, the summation of solutions is $\sum\limits_{i \in {\psi _k}} {{b_{i,k}}} $.}. We can see that the summation of the solutions increases initially, and then when it is more than 1, it is adjusted to decrease down to 1. This is because for group $\forall k \in \mathcal{K}$,  the KKT conditions indicate that the equation $\sum\limits_{i \in {\psi _k}} {{b_{i,k}}} =1$ must hold for the optimality.  By Fig. 4(c), we know the objective function, i.e., the total delay of CAVs can be reduced with the iterations and reach the minimum value in about the 20-th iteration.

\begin{table}\nonumber
	\caption*{TABLE II: Average delay and solving time}
	\vspace{-11pt}
	\begin{center}
		\begin{tabular}{c|c|c|c}
			\hline
			& {Joint-Exh-Opt} & {Exh-Opt} & {Proposed}  \\ 
			\hline
			{$t_1$ Average delay (ms)} & {69.26} & {74.47} & {80.53} \\
			\hline
			{$t_2$ Average delay (ms)} & {86.59} & {92.75} & {102.72} \\
			\hline
			{$t_3$ Average delay (ms)} & {75.18} & {80.40} & {90.44} \\
			\hline
			{$t_1$ Solving time (s)} & {9470.19} & {368.47} & {0.013}\\	
			\hline
			{$t_2$ Solving time (s)} & {9477.23} & {373.56} & {0.012}\\	
			\hline
			{$t_3$ Solving time (s)} & {9458.51} & {359.64} &{ 0.012}\\	
			\hline
		\end{tabular}
	\end{center}
	\vspace{-19pt}
\end{table}
Note that the caching programming is intended for future benefits. Since future network states are unpredictable, we are unable to plan multi-slot caching decisions by using general methods, such as the exhaustive search,  so the optimal caching is difficult to be obtained. To show the optimality of the proposed solution, we try to fix the caching decision to decouple the proposed problem into several single-slot problems. In TABLE II, we provide the comparisons in terms of CAV's delay and solving time of three schemes in three consecutive time slots under the given caching decision. The first scheme (‘Joint-Exh-Opt') provides the optimal value, which is jointly solved combining the exhaustive search for the offloading decision with the optimal resource allocation method proposed in Section IV. The second scheme (‘Exh-Opt') solves the problem in two steps, where in the first step we utilize exhaustive search to find the optimal offloading decision with the equal resource allocation, and in the second step we find the optimal resource allocation. The third scheme is the proposed scheme, consisting of two steps, where in the first step we use the trained VDN model to obtain the offloading decision, and in the second step we find the optimal resource allocation. We can see that the proposed scheme achieves good performance at an acceptable optimality gap with ‘Joint-Exh-Opt' and outperforms other schemes in terms of solving time. Considering optimality and solving efficiency, the proposed scheme is suitable for the proposed system model in our work, while ‘Joint-Exh-Opt' and ‘Exh-Opt' are not suitable due to the prohibitive solving time.   

%In order to evaluate the simulation results of figures 5, 6 and 7, we list the comparison cases:
%\begin{enumerate}
%\item woHAPS-cache0. There is no HAPS computing, which means task can be computed at local and RSU side, and the RSU side does not have caching function.
%\item woHAPS-cache100. There is no HAS computing, and the RSU has storage with a space of 100Mb.
%\item woHAPS-cache200. There is no HAPS computing, and the RSU has storage with a space of 200Mb.
%\item HAPS-cache0. This is a Local-RSU-HAPS computing system, where the RSU side does not have caching function.
%\item HAPS-cache200. This is a Local-RSU-HAPS computing system, where the RSU has storage with a space of 100Mb.
%\item HAPS-cache200. This is a Local-RSU-HAPS computing system, where has storage with a space of 200Mb.
%\end{enumerate}

 \begin{figure}[!t]
	\centerline{\includegraphics[width=7cm]{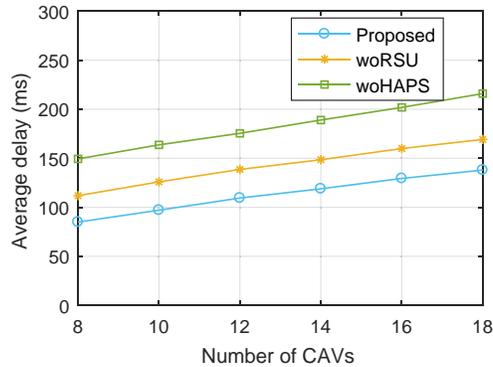}}
	\caption{Average delay vs. number of CAVs.}
\end{figure}
\begin{figure}[!t]
	\centerline{\includegraphics[width=7cm]{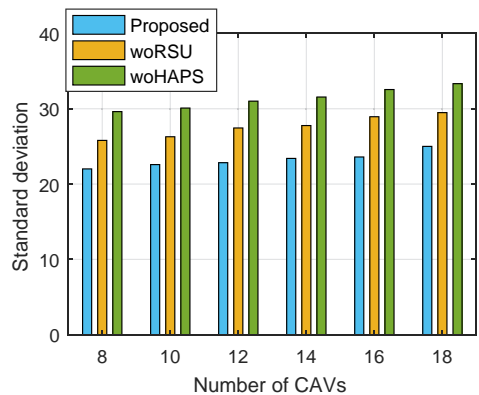}}
	\caption{Standard deviation of delay vs. number of CAVs.}
\end{figure}
 Fig. 5 shows the average delay results of the three schemes versus different numbers of CAVs. We let label  `Proposed' represent the proposed three-layer computing scheme that integrates three computing ways: local computing, RSU computing, and HAPS computing, while label `woRSU' and label `woHAPS' represent the two-layer computing schemes where there is no RSU computing and there is no HAPS computing, respectively. It can be seen that the proposed scheme always outperforms other schemes, which demonstrates that the integration of the computing resources of CAV’s onboard device, RSU, and HAPS  can effectively improve the delay performance. Comparing ‘Proposed’ and `woRSU', we can see the importance of RSU computing because RSUs are close to CAVs. When there is no RSU computing, a large number of tasks need to be offloaded to the HAPS, which leads to large communication delays for transmitting individual input data. Accordingly, it can be reflected that HAPS computing is better to be integrated with terrestrial edge computing rather than to be replaced with it. Comparing `woRSU' and `woHAPS', we know that HAPS can mitigate delays compared with RSUs. This happens due to two reasons: first, because of its large payload, the HAPS can be equipped with a more powerful computing server than RSUs, which can enhance the computational capability of the whole system. Second, HAPS computing can help to mitigate communication delays, because if some of the tasks can be computed at the HAPS, the requested fundamental data can be obtained from the HAPS library directly without any transmissions.

 Fig. 6 shows the standard deviation of delay of three schemes versus different numbers of CAVs. As can be seen that under different numbers of CAVs, `Proposed' can always achieve a smaller standard deviation value than other schemes, which means under the proposed scheme, CAVs can achieve relatively concentrated and stable delays. Besides, as the number of CAVs increases, the standard deviation value increases slightly.
%and among these cases, the mean values of the average delay are 70, 95, and 110 miliseconds.
\begin{figure}[!t]
	%\centerline{\includegraphics[width=\columnwidth]{s.pdf}}
	%\centerline{\includegraphics[width=7cm]{cdf8.eps}}
	\centerline{\includegraphics[width=8cm]{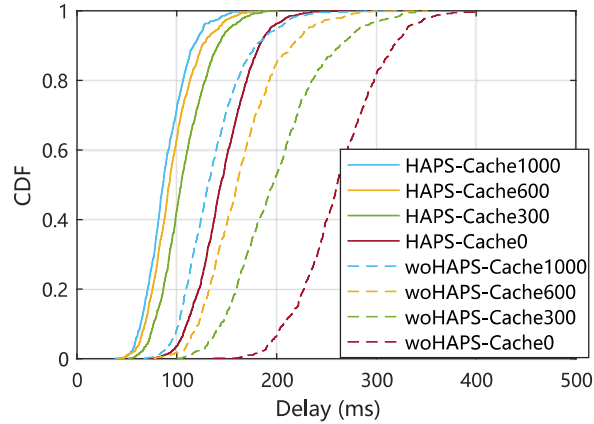}}
	\caption{CDF of delay vs. with/without HAPS computing and caching space of RSU.}
	\label{fig4}
\end{figure}

 Fig. 7  evaluates the impact of HAPS computing and the caching space on the delay performance by comparing the cumulative distribution function (CDF) of the delays. 
We let the labels ‘HAPS’ and ‘woHAPS’ indicate whether there is HAPS computing. The labels `Cache0', `Cache300', `Cache600' and `Cache1000' indicate an RSU caching space of 0 Mbits (i.e., no caching at RSU), 300 Mbits,  600 Mbits, and 1,000 Mbits, respectively. 
 As we can see, the cases with HAPS computing achieve low delays at higher probabilities than the cases without HAPS computing, which can be explained by the same two reasons for the delay mitigation by HAPS computing as mentioned in the description of Fig. 5. Also, we can see that the cases without caching are the worst, and the larger the caching space, the higher the probability of achieving low delay performance. This is because when there is no caching at RSUs, the fundamental data will be delivered from the HAPS to the ground if local or RSU computing is selected, which definitely results in long-distance communication delays. By increasing the caching space, more fundamental data can be stored in advance, and therefore when doing local or RSU computing, more fundamental data can be obtained without experiencing long-distance transmission from the HAPS. To further verify the above-mentioned conclusions, we provide the communication and computation delays in Fig. 8 and Fig. 9, respectively. 
 \begin{figure}[!t]
 	%\centerline{\includegraphics[width=\columnwidth]{s.pdf}}
 	%\centerline{\includegraphics[width=7cm]{figure4a_commu_delay_bar.eps}}
 	\centerline{\includegraphics[width=11cm]{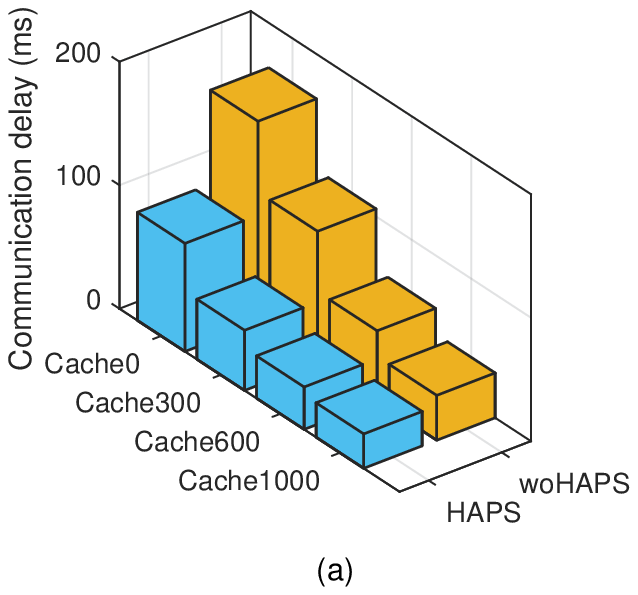}}
 	%\centerline{\includegraphics[width=10cm]{figure4b_commu_opt_equ.pdf}}
 	\centerline{\includegraphics[width=11cm]{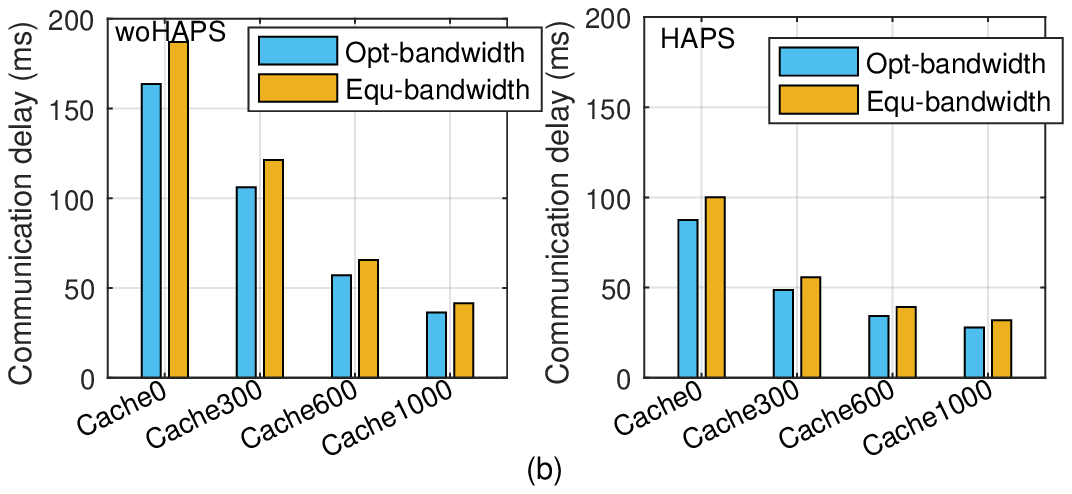}}
 	\caption{(a) Communication delay vs. caching space and with/without HAPS computing. (b) The comparison of the optimal bandwidth allocation and the equal bandwidth allocation.}
 \end{figure}

%because computing task at HAPS can help to mitigate not only the computation delay performance but also the  when tasks can be computed at the HAPS, the requested fundamental data can be obtained directly, thus improving the communication delay

%For example, the cases `HAPS-woCache', `HAPS-Cache100' and `HAPS-Cache200' obtain a delay of less than 0.5 s  at probabilities of 0.02, 0.47 and 0.8, respectively.

%For instance, between the two cases `HAPS-Cache200' and `woHAPS-Cache200', the former can achieve a delay of less than 0.5 s with a probability of 0.8, and the latter only with a probability of 0.18, b

Fig. 8 evaluates the communication delay performance.  Fig. 8(a) shows the impact of HAPS computing and the caching space on the communication delay performance. Comparing the cases ‘Cache0’, ‘Cache300’, ‘Cache600’, and ‘Cache1000’,  we see that the introduction of caching at RSUs can reduce the communication delays significantly, and as more  fundamental data is stored at RSUs, the communication delays can be further mitigated.  Comparing the cases ‘HAPS’ and `woHAPS', we can see that the introduction of HAPS computing can reduce the communication delay. %Especially, the introduction of HAPS can mitigate the communication delays significantly  for the case `Cache0'. This is because when there is no caching at RSUs, `woHAPS' means that all the requested fundamental data must be transmitted from the HAPS to the ground because of the requirements of local and RSU computing; however, HAPS computing can fetch data directly from the library, so that the large communication delays caused by long-distance transmission can be avoided.
% Therefore, if we introduce HAPS computing in this case, the communication delays will drop obviously. 
Fig. 8(b) shows a performance comparison of communication delays with different bandwidth allocation methods, where ‘Opt-bandwidth’ and ‘Equ-bandwidth’ indicate the optimal method proposed in Section IV and equal method for allocating bandwidth, respectively. As we can see, the optimal bandwidth allocation can effectively reduce communication delays.

\begin{figure}[!t]
	%\centerline{\includegraphics[width=\columnwidth]{s.pdf}}
	\centerline{\includegraphics[width=11cm]{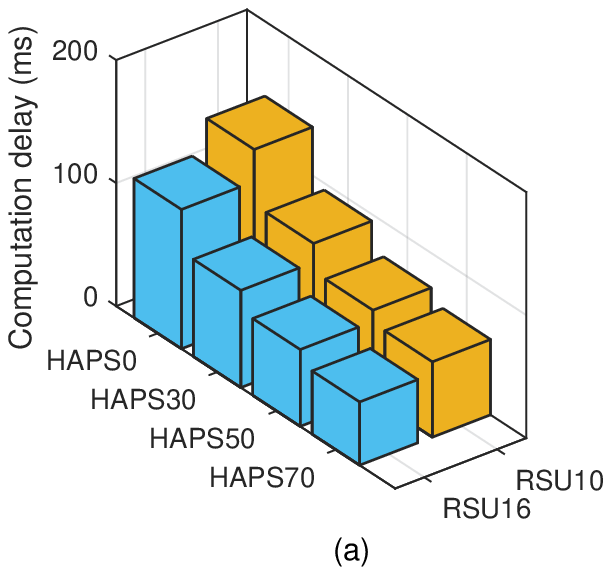}}
	\centerline{\includegraphics[width=11cm]{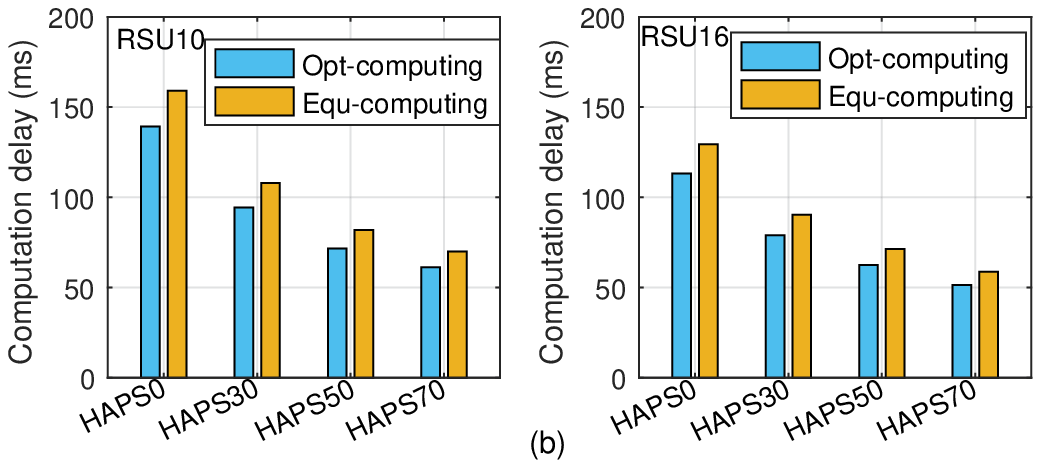}}
	\caption{(a) Computation delay vs. computational capabilities of HAPS and RSU. (b) The comparison of the optimal computing resource allocation and the equal computing resource allocation.}
\end{figure}

%This is because when the frequently requesting fundamental data can be stored, the long distance transmission of large amount of data can be avoided. 
Fig. 9 evaluates the computation delay performance. Fig. 9(a) shows the impact of the computational capabilities of the HAPS and RSUs on the computation delay performance.  We let labels ‘HAPS0’, ‘HAPS30’, ‘HAPS50’, and ‘HAPS70’ indicate the cases where the computational capability of the HAPS is 0 G, 30 G, 50 G, and 70 G in CPU cycle/s, respectively. The labels `RSU10' and `RSU16' indicate the cases where the computational capabilities of RSUs are 10 G and 16 G in CPU cycle/s. Comparing ‘HAPS0’, ‘HAPS30’, ‘HAPS50’, and ‘HAPS70’, we can observe that the introduction of HAPS computing can reduce the computation delays, and by increasing the computational capability of the HAPS, the computation delays can be mitigated further since more computing resources of the HAPS are complemented to terrestrial networks.  By comparing ‘RSU10’ and ‘RSU16’, two points may be observed. First, the impact of the computational capability of RSUs on computation delays is evident because the RSUs are physically closer to CAVs and play an important role in computing. Second, when the computational capabilities of RSUs are lower, HAPS computing can significantly mitigate computation delays. This indicates that the introduction of HAPS computing is even more important for areas with severe  resource shortages, such as remote and congested areas. In Fig. 9(b), we provide a comparison of computation delay performance for different computing resource allocation methods, where ‘Opt-computing’ and ‘Equ-computing’ indicate the optimal method proposed in Section IV and equal method for allocating computing resources. We can see that Fig. 9(b) verifies the superiority of the optimal computing resource allocation.

\begin{figure}[!t]
	%\centerline{\includegraphics[width=\columnwidth]{s.pdf}}
	\centerline{\includegraphics[width=8cm]{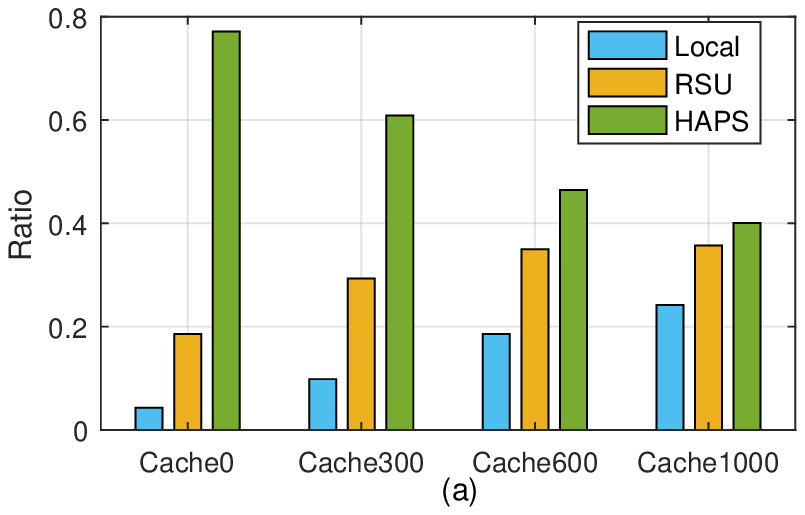}}
	\centerline{\includegraphics[width=8cm]{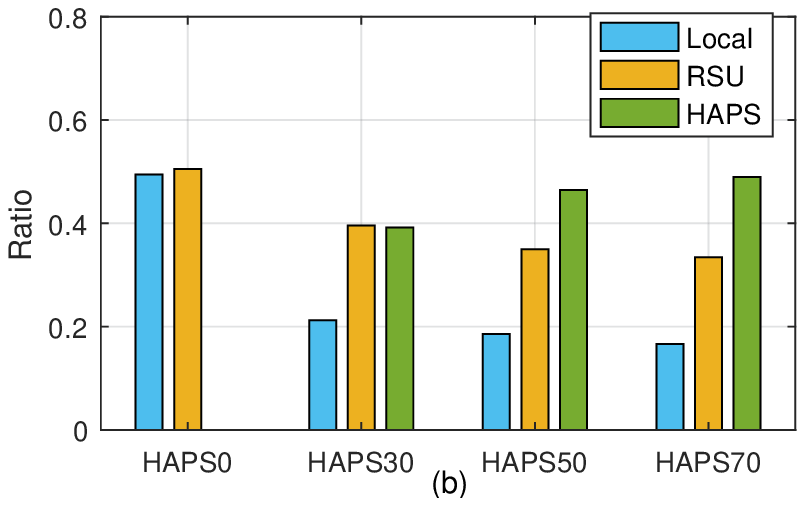}}
	\caption{(a) Computing ratio vs. caching space of RSU. (b) Computing ratio vs. computational capability of the HAPS.}
	\label{fig6}
\end{figure}
Fig. 10 shows the  ratio of three computing ways: local, RSU, and HAPS computing. Fig. 10(a) shows the impact of the caching space of RSUs on the computing ratio. We can see that HAPS computing is the main computing way, especially when there is no caching, with a computing ratio of 78\%. This is because when there is no caching at RSUs, more CAVs select HAPS computing and fetch the requested fundamental data from the HAPS library directly so that the long-distance transmission can be avoided. By increasing the caching space, the ratio of HAPS computing decreases, while the ratios of local  and RSU computing increase. The reason is that as more contents can be stored at RSUs, the requested fundamental data is more likely to be obtained from the close RSUs, thereby reducing the communication delays and increasing local and RSU computing ratios. Fig. 10(b) shows the impact of the computational capability of the HAPS on the computing ratio. We can see that, as the computational capability of the HAPS increases, HAPS computing ratio increases, and the local computing ratio decreases. This phenomenon shows that the increase of HAPS's computational capability makes HAPS computing play an increasingly important role in the system.     
\begin{figure}[!t]
\centerline{\includegraphics[width=11cm]{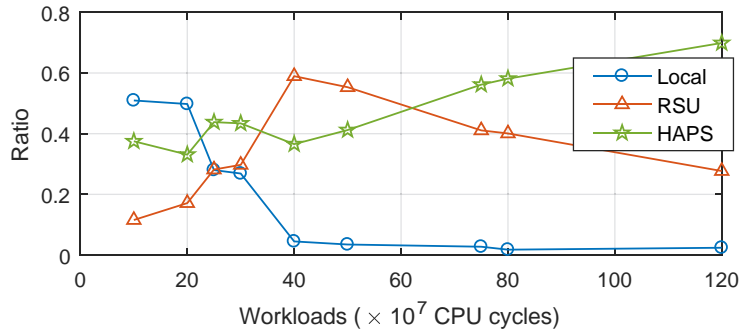}}
\caption{Computing ratio vs. workloads.}
\label{fig7}
\end{figure}

Fig. 11 evaluates the ratio of three computing ways for processing the tasks with different workloads. We can see that local computing tends to process tasks with small workloads because the computational capability of each CAV is small.  HAPS computing tends to process tasks with large workloads since these tasks require powerful computational capabilities. As for RSU computing, it can help to process more tasks with medium workloads. Therefore, we know that three computing ways play different roles for processing the tasks with different workloads. 

\begin{figure}[!t]
	\centerline{\includegraphics[width=7cm]{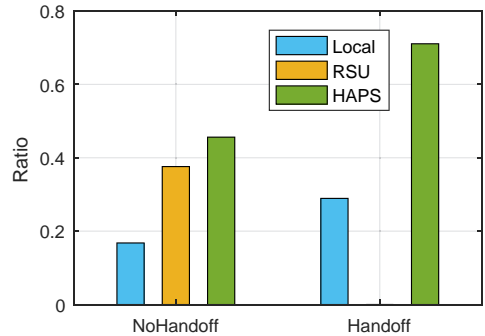}}
	\caption{Computing ratio vs. no handoff and handoff situations.}
\end{figure}
To illustrate the impact of CAV's mobility on decision-making, we make a simple extension of our work. Specifically, we set a communication range for the RSU as 200 m. When the CAV exceeds this range, it will trigger a handoff with the RSU, which means that the CAV cannot offload data to RSU until it can enter the next one.
 In Fig. 12, we provide the ratio of three computing ways for two situations, i.e., no handoff occurs and handoff occurs, indicated by `NoHandoff' and `Handoff', respectively. We can see that in the absence of handoff, the ratio of choosing HAPS computing is about 45\%, and when the network handoff occurs, this ratio has increased to more than 70\%. This demonstrates that HAPS plays a more important role in the handoff situation.

\section{Conclusion}
In this article, we studied a caching and computation offloading scheme with the assistance of HAPS to improve delay performance at ITS. Specifically, this scheme formulated a local-RSU-HAPS computation framework by integrating the HAPS with terrestrial computing networks, where the HAPS had a library of fundamental data. In addition, the caching technique was used at network edges to further mitigate the large transmission delays for delivering the requested fundamental data when executing ITS-based tasks. We formulated the problem as a delay minimization problem, where we first focused on optimizing the computation offloading and caching decisions and then focused on the allocations for bandwidth and computing resources. By utilizing the multi-agent reinforcement learning and Lagrangian methods, we solved the proposed problem.  The simulation results demonstrated the advantages of the proposed paradigm in improving delay performance. Although the delay performance with  a HAPS-assisted ITS has been investigated in this paper, energy efficiency was not discussed, which is important for the HAPS network. This topic will be studied in our future work.

\ifCLASSOPTIONcaptionsoff
  \newpage
\fi
\small
\bibliographystyle{ieeetr}
\bibliography{IEEEabrv,ICCworkshop}

\end{document}